\documentclass[letterpaper]{article}
\usepackage{natbib,alifeconf}
\usepackage{hyperref}
\usepackage{amsmath}

\title{Evolutionary rates of information gain and decay in fluctuating environments}
\author{Nicholas Guttenberg$^{1}$
\mbox{}\\
$^1$Earth-Life Science Institute, Tokyo Institute of Technology, Tokyo, Japan  \\
ngutten@gmail.com}

\begin{document}
\maketitle

\begin{abstract}
In this paper, we wish to investigate the dynamics of information transfer in evolutionary dynamics. We use information theoretic tools to track how much information an evolving population has obtained and managed to retain about different environments that it is exposed to. By understanding the dynamics of information gain and loss in a static environment, we predict how that same evolutionary system would behave when the environment is fluctuating. Specifically, we anticipate a cross-over between the regime in which fluctuations improve the ability of the evolutionary system to capture environmental information and the regime in which the fluctuations inhibit it, governed by a cross-over in the timescales of information gain and decay.
\end{abstract}

Intuitively, evolution is a process by which populations learn about the world. As such, information appears to be a natural concept for constructing an abstracted view of evolutionary process --- how much information does the population have about the environment, how much information is retained between generations, how much more information is needed before the population can find a fitness optimum, and how much of that per generation does selection manage to bring into the population? Along the lines of \citep{adami2004information}, these ideas can be connected to specific information theoretic quantities and bounds. However, while such information theoretic quantities can be measured after the fact, we would like to see whether a formulation of evolutionary dynamics in terms of information is sufficiently resolved to be predictive of the behavior of that evolving system in different conditions. If we were to reduce the details of an evolving system's state to a set of information quantities, could we find equations of motion purely in terms of those variables which would still predict the future dynamics of that evolutionary process to some degree?

In this paper, we target the response of evolving systems to fluctuating environments in order to investigate this idea. Specifically, can we predict when fluctuations would help or hinder a population's ability to adapt, simply by looking at the dynamics of the information between population and environment? The literature on fluctuating environments has examples of both cases. In terms of positive contributions, they can accelerate adaptation and even increase the asymptotically achievable fitness \citep{kashtan2007varying}. It has also been proposed that the richness of open-ended evolution may depend on (or even originate from) the correspondingly richer problems posed by needed to be well-adapted to a multiplicity of environments \citep{neumann1997organizational, bedau2000open}. Environmental variations have also been considered as a driving source behind  multi-level organization and generalization properties of evolutionary systems \citep{medernach2017comparative}. At the same time, it is observed that environmental variations may induce evolving systems to make tradeoffs between optimality and robustness \citep{levins1967theory, wilke2001evolution}, and adaptation between a succession of sufficiently unrelated environments can interfere with adaptation \citep{steiner2012environmental}.

In order to try to understand these tradeoffs, we consider a simplified case in which there are two independent environments given by random variables $E_1$ and $E_2$, such that the population of organisms ${\vec{g}}$ can have some mutual information with them $I({g};E_1)$ and $I({g};E_2)$. Here, we further specialize to the case in which these environments have no mutual information with each-other $I(E_1;E_2) = 0$, and that mutation operates independently from the environmental random variables. Based on the constraint that the only way for $I({g};E)$ to increase is through selection, we can construct a simple model in which while the population is being exposed to $E_1$, $I({g};E_2)$ only has decreasing terms, and vice versa. If we then average over a cycle including both $E_1$ and $E_2$, we can consider when the gain of mutual information during the selection phase would be balanced against the loss of information during the neutral phase. For a rapidly varying environment, this balance point should only depend on the instantaneous rates of change of the mutual information around the population's steady state, whereas for slowly varying environments, the overall shape of the trajectory may lead to systematic variation in things like the rates of increase and decrease.

We make a further assumption --- that the underlying mechanisms responsible for determining the rate of increase and decrease of the mutual informations are intrinsic to the dynamics of replication, mutation, and selection and should be the same both in slowly varying and quickly varying environments. If this is the case, then the balance point between information gain and loss in the rapidly fluctuating environment can be related to observations of the same evolutionary system made when the environments vary at a different rate. However, if for example the fluctuations led to significantly different population structures in steady state, then that would violate this assumption. We will ultimately return to this point, as it is likely that this does in fact occur, and may be connected to the relationship between fluctuating environments and the evolution of evolvability \citep{ofria2016evolution}.

\section{Related Work}

The sense in which evolved organisms contain information about the world has been explored in a variety of ways. One approach centers on looking at the Shannon information, concerned primarily with measures of the entropy of different parts of the genome \citep{schneider2000evolution, chang2005shannon}. This naturally extends into methods which use internal mutual informations between different bases, protein residues, and structures as a way of understanding things such as co-evolution and structured variation \citep{gobel1994correlated, martin2005using, gloor2005mutual}. These studies often relate to understanding the dynamics of neutral evolution. On the other hand, there is the question of what if anything a given portion of genetic entropy is 'about' --- that is to say, not just whether it varies, but whether it has information about variations that exist within the environment or context of the organism \citep{adami2004information}. While identifying the random variables which characterize a given environment in nature is in general ambiguous, cases of co-evolution between competing sets of sequences can make this concrete by treating the sequences of one population (or corresponding phenotypic variations) as the environment of the other population and vice versa. For example, \citep{xia2009using} looks at mutual information in the co-evolutionary dynamics between a virus and the immune response as a way to understand immune escape.

\section{Evolutionary simulations}

We consider a simple evolutionary simulation where each organism consists of a binary string $\vec{g}$ of length $L$ ($L=50$ for all simulations presented in this paper), and each environment consists of a pair of binary strings: the target sequence $\vec{E}$, and the mask $\vec{m}$. The elements of $\vec{E}$ are randomly $0$ or $1$ with equal probabilities, whereas the elements of $\vec{m}$ are $1$ with probability $\Gamma$ and $0$ otherwise, where $\Gamma$ measures the fraction of the organism's capacity a given environment can be expected to take up. The relative fitness of each organism in the population is then given by:

\begin{equation}
F = 1 + \sum_i^L \vec{m}_i \left( 1-| \vec{g}_i - \vec{E}_i | \right) 
\label{Fitness}
\end{equation}

This means that for a given environment there are a subset of specific sites (given by $m_i$) which have non-neutral fitness effects, and each site has an independent preferred value of the genome given by $\vec{E}_i$. The fitness is then linear in the number of sites which are matched. The independence between sites is chosen in order to enable a simplified sense of the mutual information to be used, so that we can factorize the overall joint distribution over sequences into a product of distributions at each site. Each generation, fitnesses are evaluated and a new population is composed by randomly sampling from the old population with replacement in proportion to fitness, so that the population size remains constant ($N=200$ in all simulations reported in this work). Mutation occurs with a per-base probability $\mu/L$.

In this system, we wish to consider the mutual information between the population of sequences and the environment, and how it changes over time. The mutual information between random variables $x$ and $y$ is defined:

\begin{equation}
 I(x;y) \equiv \int p(x,y) \log \left( \frac{p(x,y)}{p(x)p(y)} \right) dx dy
\end{equation}

This quantity places a bound on the ability to better infer $x$ given an observation of $y$ compared to not having that observation of $y$ --- that is to say, if one knows $p(x)$ and then observes $y$, the mutual information measures the change in the entropy in the possibilities that $x$ may take conditioned on knowing $y$. The quantity is symmetric, such that the same would be true for inferring $y$ by predicting $x$. In the context of organisms and environments, the mutual information is not a quantity which would be defined in a single evolutionary trajectory but instead is a statistical property over entire ensembles of trajectories associated with a distribution over the different environments. We cannot ask 'what is the mutual information between this one organism and its environment?', but we can ask 'what is the mutual information between organisms and their environments in this evolutionary context?'

Measuring the mutual information directly in practice requires accumulating sufficient observations in order to construct estimates of $p(x,y)$, $p(x)$, and $p(y)$ via sampling. If the sum of the dimensions of $x$ and $y$ is large, this becomes prohibitively expensive to brute force as the number of samples required grows exponentially in the dimensions of the variables involved. However, due to the relationship between the mutual information and the bound on what can be inferred about the variables by observing each-other, any method of approximate inference can be used to generate a lower bound on the mutual information between variables. For example, there are techniques to use neural networks to estimate mutual informations between high dimensional random variables \citep{belghazi2018mine}. These methods could be used to extend this type of analysis to systems with arbitrary genotype-phenotype maps including the effects of epistasis, or even to systems where the information-carrying degrees of freedom are not a priori known such as chemical reaction networks.

Because our simulations consider only fitness landscapes without epistasis, we can consider the mutual information independently on per-base basis. That is to say, the relationship between each base in the genome and the corresponding base of the environmental random variable in the fitness function is statistically independent from all of the other bases, and each base only interacts with the corresponding base of the environmental random variable. Any interactions between bases occur only through aggregation into the fitness, and such interactions are independent of the specific environment sequences --- that is to say, by observing the aggregate fitness, we would receive zero information as to the value of any particular base in the environment string. As such, the total mutual information between genome and environment is just the sum of per-base mutual informations, and we can reduce the multidimensional mutual information estimation problem into a collection of independent one-dimensional estimations for which direct sampling is sufficient.

\section{Results}

We measure the per-base mutual information by sampling over $5000$ runs of the evolutionary simulation with independent random choices for the environments in each run. In order to sample the probabilities $p(x,y)$, $p(x)$, and $p(y)$ for a given base $i$, we first take the subset of runs in which that base is not masked out in the target environment. Then, from that set of runs, we measure the fraction of the population in each run which contains a $1$ at that site $\bar{g}_i = \langle \vec{g}_i \rangle$. This results in approximately $5000 \Gamma$ scalar values for each base. These values are quantized into a histogram with $100$ bins between $[0,1]$, and we then accumulate samples from the unmasked runs to estimate $p(\bar{g}_i, \vec{E}_i)$, $p(\bar{g}_i)$, and $p(\vec{E}_i)$. With these discrete distributions, we can directly evaluate the mutual information between population and environment as a function of time.

A run of the simulations involves first letting the population adjust to a 'burn in' environment for $200$ generations where data are not taken, followed by $200$ generations in environment $E_1$ and $200$ generations in environment $E_2$ (the mask vectors also vary between these environments). We then measure the informations $I(\bar{g}(t); E_1)$ and $I(\bar{g}(t); E_2)$ per (coding) base. In the 'varying environment' experiments, we switch environments between $E_1$ and $E_2$ every generation, for the same total of $400$ generations. Code for these simulations, results, and subsequent analysis are available at \url{https://github.com/ngutten/evolution_infodynamics}.

\begin{figure}
\includegraphics[width=\columnwidth]{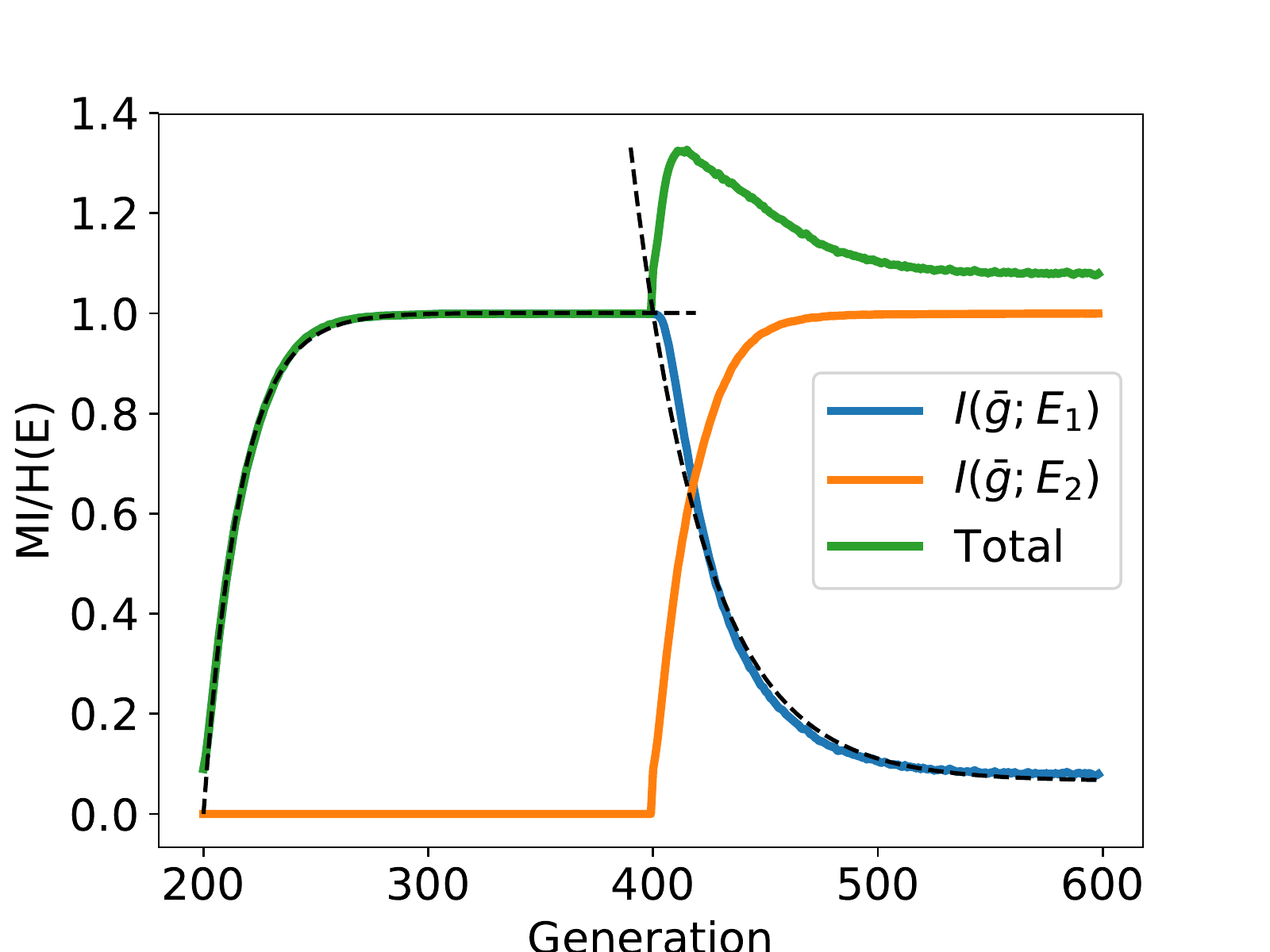}
\caption{\label{InfoDynamics} Dynamics of mutual information between population and environment for $\mu = 0.5$ and $\Gamma = 0.2$. Information measures are normalized by the maximum information per base associated with a single environment for that value of $\Gamma$. Dashed lines are fits to Eq.~\ref{SimpleDynamics} }
\end{figure}

The first set of results involves the dynamics of the mutual information with respect to time, and are shown in Fig.~\ref{InfoDynamics}. The observed dynamics of the mutual information under selection are quite close to a saturating function of the form $I(t) = A (1-\exp(-t/\lambda_+))$. Meanwhile, the decay behavior when a different environment is being selected for appears to closely follow the form $I(t) = A \exp(-t/\lambda_-)$.

This type of saturating behavior is consistent with dynamics in which there is a constant source and a linear decay:

\begin{equation}
\partial_t I_i = 
\begin{cases} 
-I_i/\lambda_+ + A/\lambda_+, & \textrm{Selection on } E_i \\
-I_i/\lambda_-, & \textrm{Selection on } E_j \\
\end{cases}
\label{SimpleDynamics}
\end{equation}

\begin{figure}
\includegraphics[width=\columnwidth]{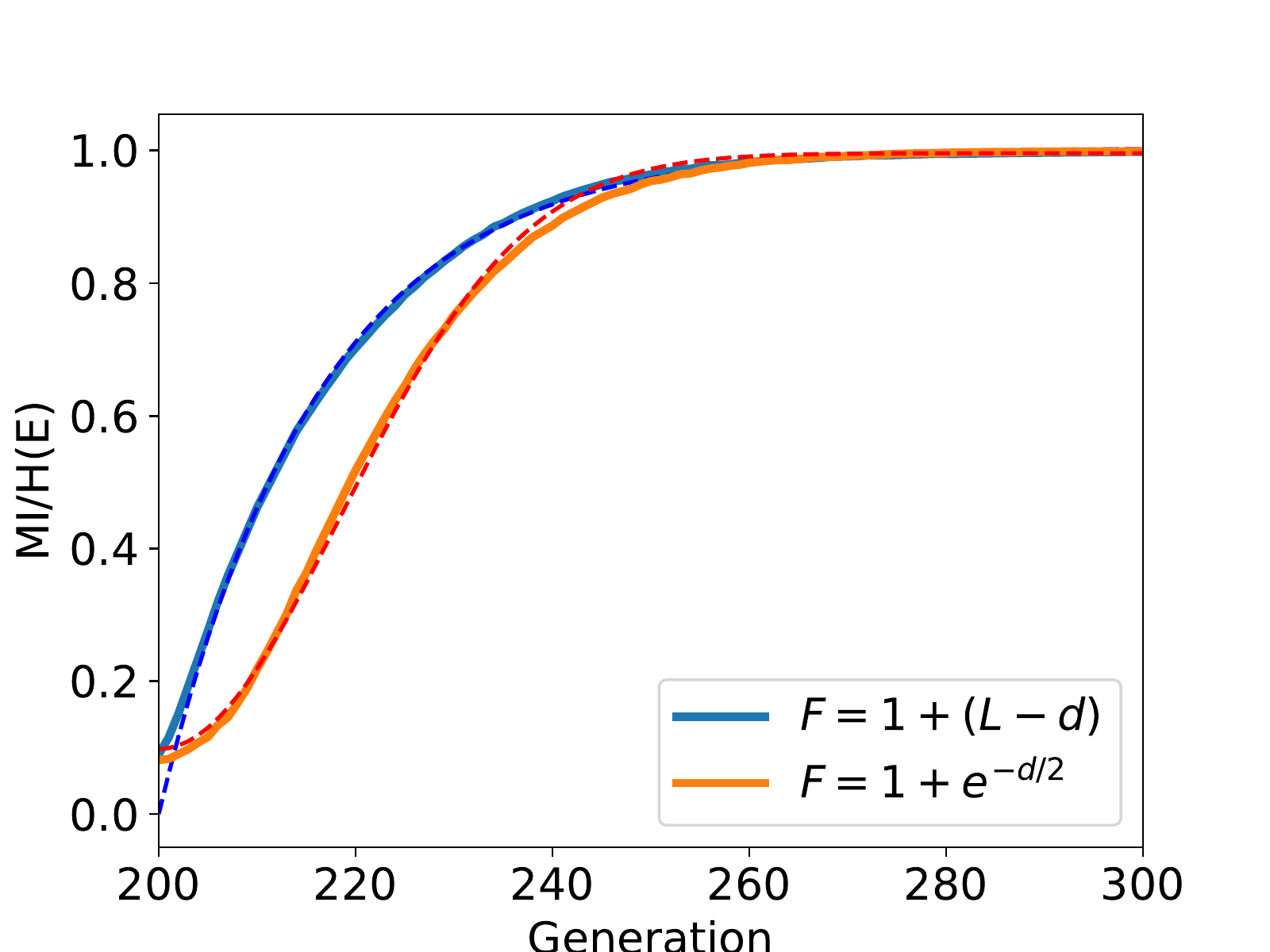}
\caption{\label{DynamicsForm} Comparison between the initial increase of mutual information for a linear fitness function versus an exponential fitness function. The blue solid line corresponds to a linear fitness function as given by Eq.~\ref{Fitness}, and the blue dashed line is a sigmoidal fit. The red solid line corresponds to a fitness $F=1+\exp(-d/2)$ where $d$ is the Hamming distance between target and environment, and the red dashed line is a fit to $A (1-\exp(-t^2/\lambda_+^2))+B$. Parameters are $\mu=0.5$ and $\Gamma=0.2$.}
\end{figure}

The fact that $\lambda_-$ and $\lambda_+$ are not the same suggests that the decay is not just from mutation (which should be the same in both cases), but rather includes an effect where selection for one environment can influence the rate at which mutual information with a second environment is lost. It is a bit surprising that the form would be this simple, as why would selection provide information at a constant rate rather than one which depends on how the population is positioned relative to the fitness landscape? It may just be that for our particular fitness function, since each move towards the fitness optimum provides the same selective contrast regardless of how close or far an organism is to the optimum, that these effects are zero. In fact, if we switch to a fitness landscape in which the fitness is exponential in the Hamming distance from the optimum, we see kinetics of the form $A(1-\exp(-t^2/\lambda_+^2))+B$ instead (Fig.~\ref{DynamicsForm}), so this is not universal.

Given that we do observe this form of kinetics over a large range of parameters for our case, we can attempt to use Eq.~\ref{SimpleDynamics} to relate the kinetics of saturating adaptation in the case of one environment to what would happen in a system which oscillates between two environments with period $T$, by balancing the information gained during the selection phase against the information lost during the non-selected phase:

\begin{equation}
AT/\lambda_+ - \int_0^T I(t)/\lambda_+ dt - \int_T^{2T} I(t)/\lambda_- dt = 0
\end{equation}

For rapidly oscillating environments ($T\rightarrow 0$), this criterion is satisfied when:

\begin{equation}
I = A \frac{ 1 }{ 1 + \lambda_+ / \lambda_-}
\label{Criterion}
\end{equation}

This is in comparison to an asymptotically slowly-varying environment, in which the information about one environmental random variable $I_1 \rightarrow A$, while the other $I_2 \rightarrow 0$. In this case, since the different environments are independent ($I(E_1; E_2)=0$), the total mutual information between the population and the set of environments is additive $I({g}; {E_1, E_2}) = I({g}; E_1) + I({g}; E_2)$. We can now ask, at least in the context of this model, when does a fluctuating environment result in the population having more information about the set of environments in total than if it just adapted to a single environment? In the context of the criterion given by Eq.~\ref{Criterion}, we should expect this to happen when the timescales of information gain and decay are equal $\lambda_+ = \lambda_-$ because at that point, each environment contributes half of its total entropy to the total information ($I_1 = I_2 = A/2$), and so the total information the system has about the oscillating environment pair is equal to the total information the system would have about one single static environment.

In the example of Fig.~\ref{SimpleDynamics}, we see that there is a point at which the sum total information between the system and both environments exceeds the entropy of a single environment --- the information content is greater than what could be obtained if an environmental switch was not present. Similarly, we expect there to be cases in which switching between environments prevents the entirety of the available entropy of a single environment to be transferred into the system. According to our simplified model, we predict that the cross-over between these cases should occur when the timescales of information gain and decay are equal. As such, we will test the simplified model by empirically measuring those timescales, and then comparing the cross-over in timescales to the point at which the excess information peak disappears.

\subsection{Measuring timescales}

\begin{figure*}
\includegraphics[width=\textwidth]{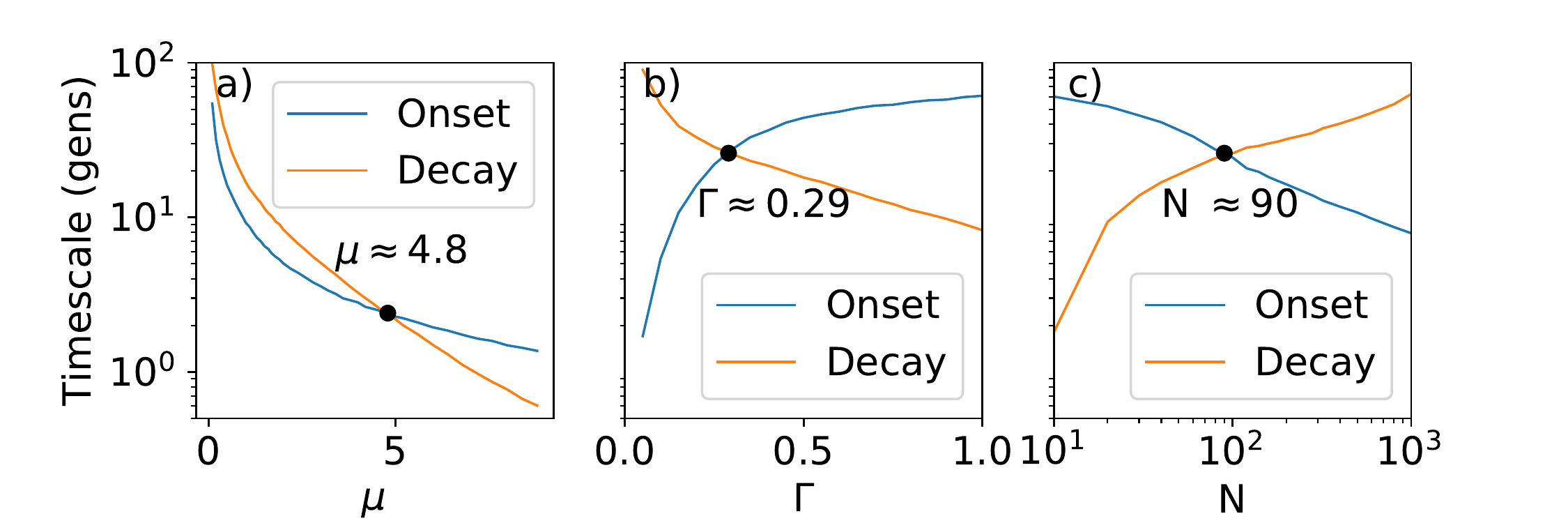}
\caption{\label{ParamPlot} Variation of the time constant of information gain and loss with $\mu$ (a), $\Gamma$ (b), and population size $N$ (c). The baseline parameters are $\Gamma=0.2$, $\mu=0.5$, and $N=200$, with the three panels each corresponding to varying one quantity around that baseline. }
\end{figure*}

In order to test this relationship between the timescales of information gain and decay, we fit the observed dynamics of mutual informations in the slowly varying environments to Eq.~\ref{SimpleDynamics} and look at how the timescales $\lambda_+$ and $\lambda_-$ vary with $\mu$ (Fig.~\ref{ParamPlot}a), $\Gamma$ (Fig.~\ref{ParamPlot}b), and population size (Fig.~\ref{ParamPlot}c). Unsurprisingly, as we increase the mutation rate, information is lost from the system more rapidly. However, increasing the mutation rate also increases the rate at which the information held by the system about the environment approaches it's asymptotic value. Increased mutation rate may ultimately lead to less information being retained --- and so in terms of total information gain by some fixed point in time $\tau$ we would expect there to be a local maximum with respect to mutation rate. However, when that total information gain is normalized out, the remaining effect of mutation on the timescale is monotonic.

When we inrease the fraction of the non-neutral part genome $\Gamma$, we are in effect increasing the amount of information there is to learn about a given environment, and correspondingly this means that we are decreasing the relative information capacity with respect to the genome and the size of an optimal solution for a given environment. As such, the behavior of the onset and decay phases with respect to increasing $\Gamma$ are markedly different. As increased $\Gamma$ means there is more information to learn before convergence, the slope of the fraction of the total entropy captured is proportionally decreased and so the timescale of learning is correspondingly slowed (so $\lambda_+$ becomes larger with increasing $\Gamma$). On the other hand, as the genome is closer to capacity, once the environment changes then a larger fraction of bits are in conflict and so experience decay at the rate imposed by selection effects, rather than the rate imposed by mutation effects. In response, $\lambda_-$ monotonically decreases with increasing $\Gamma$.

Population size has a strong impact on timescales as well. Even if an individual organism experiences a mutation in a particular base, there are redundant copies of that information distributed among the population. As such, information about the previous environment decays not just at the rate of an individual mutating, but instead at the rate at which that mutation would proceed towards fixation. For neutral mutations, this timescale is  linear in the population size (Eq.~14 of \cite{kimura1969average}), whereas for non-neutral mutations it is (to first order) logarithmic in the population size (Eq.~50 of \cite{uecker2011fixation}). So it makes sense that as we increase the population size, we generally see an increase in the decay timescale $\lambda_-$. At the same time, a larger population means that selection has an increased bandwidth for transferring information about the environment into the system --- and so the rate of information gain accelerates, and the onset timescale $\lambda_+$ becomes shorter. 

Since the characteristic time-scale of information gain under a new selection pressure generally differs from the characteristic time-scale of information loss about previous environments, meaning that the population can end up with either an information excess (in that it stores more adaptive information than would be necessary to maximize fitness in the current environment alone) or an information deficit (in that adaptation to the new environment causes the old one to be forgotten more quickly than new information comes in). Holding $\Gamma$ constant, we see that there is a particular value of the mutation rate at which this cross-over occurs, although higher mutation rates reduce both timescales strongly. In comparison, holding $\mu$ constant, $\Gamma$ also has a cross-over point but the effect is much stronger. 

\subsection{Varying environments}

\begin{figure*}
\includegraphics[width=\textwidth]{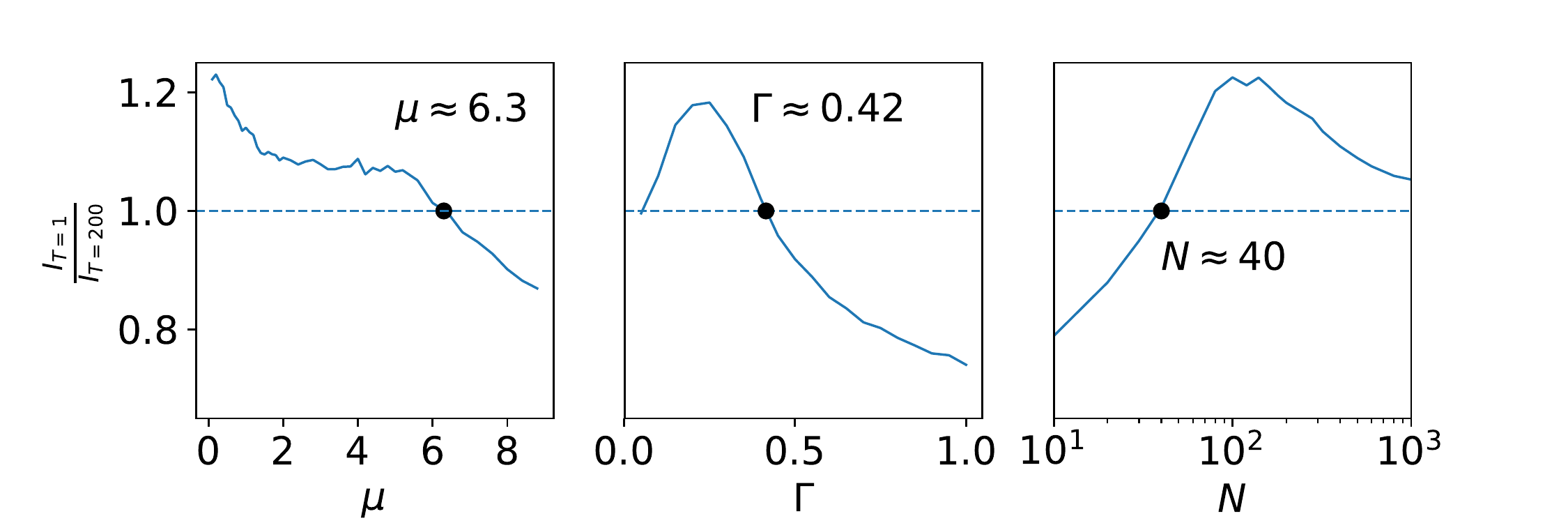}
\caption{\label{Oscillation} This figure shows the ratio of the peak sum information between a system with changes between a pair of environments every generation, and one in which a single change occurs at $T=200$. The left panel shows the case in which the mutation rate $\mu$ is varied with a constant $\Gamma = 0.2$; the center panel shows the case where $\Gamma$ is varied with a constant $\mu = 0.5$; and the right panel shows variation in population size, given constant $\Gamma = 0.2$ and $\mu = 0.5$. In the case of $\mu$ and $\Gamma$, the cross-over points occur at larger values than the corresponding cross-over between timescales in the static environment by about a factor of $\approx 1.4$: $\mu = 6.3$ versus $\mu = 4.8$ for static, and $\Gamma = 0.42$ versus $\Gamma = 0.29$ for static. For the population size effect, the cross-over occurs earlier at around $N=40$, compared to $N=90$ for equal timescales.}
\end{figure*}

This now brings us to the case of varying environments. This cross-over point indicates that, for one switching event, the effect can either be to temporarily increase the information content of the system about both environments or temporarily decrease the information content of the system below where it would otherwise be without the switch. If we now have multiple switching events on a timescale comparable to the scale of information gain and decay, we might expect the parameters of the system relative to this cross-over point to determine whether the effect of environmental variations is force extra information into the system on the net, or to force information out of the system on the net. We show the peak total information observed in environments oscillating with period $T=1$ in Fig.~\ref{Oscillation} in comparison with the same system at $T=200$ both for varying $\mu$ and for varying $\Gamma$. As expected, we see a cross-over between the case in which fluctuations drive excess information into the system and the case in which fluctuations drive information out of the system. While the cross-over points are similar to the point at which the time-scale of information gain and information decay are balanced, they do disagree in detail by a factor of around $1.4$. So while the broad idea that balance between information gain and loss informs us about how an evolutionary system would respond to a fluctuating environment, the actual relationship seems to differ in some details.

\section{Conclusions}

We have demonstrated a measurement of the dynamics of mutual information between population and environment in the case of a simplified binary sequence evolutionary simulation. The dynamics of information in this system exhibit sigmoidal growth and decay curves, consistent with the idea of a constant rate of information injection balanced against proportional information loss. Where that holds, changes in the environment result in dynamics with a characteristic time-scale associated the gain of information about the new environment and a different characteristic time-scale associated with loss of information about the old environment. Depending on mutation rate and the amount of information associated with environments relative to the genome capacity, these time-scales vary and may undergo a cross-over at particular values of the mutation rate and capacity. We expect from the sigmoidal model that this cross-over would correspond to the point at which a rapidly fluctuating environment would asymptotically induce either a net gain or loss of information in the population --- in essence, determining whether or not adapting to multiple environments would enhance or inhibit the evolutionary process. While we found such a transition, the location of the transition differed by a significant factor from the point which would be predicted by looking at the time-scales alone, suggesting that there is still some additional consideration for how fluctuations interact with the evolutionary dynamics that the simplified information flow model is missing.

One potential factor is that ability of a population to gain information from the environment depends on the population's structure with respect to the structure of the environment --- that is to say, selection does not simply introduce a constant transfer of information from environment to population, but rather causes changes in the population structure which have consequences for how much information will be able to flow from the environment to the population in subsequent generations. In both \citep{steiner2012environmental} and \citep{ofria2016evolution} a connection is made between environmental fluctuations and the evolution of evolvability. Similarly, \citep{virgo2017lineage} suggests that even for simple fitness landscapes, lineage effects can lead to significant adaptation of the evolvability of the population. These factors are not captured by the simple rate model of information flow. However, at the same time, an investigation into how the rate of information flow within an evolving system changes over time could be a useful way to probe these evolvability effects in the future.

\section*{Acknowledgements}

We would like to acknowledge Nathaniel Virgo and Harrison Smith for helpful feedback on the contents of this paper.

\bibliographystyle{apalike}
\bibliography{bibliography}

\end{document}